\documentclass[a4paper,fleqn,usenatbib]{mnras}

\usepackage{newtxtext,newtxmath}
\usepackage[T1]{fontenc}
\usepackage{ae,aecompl}

\usepackage{graphicx}	% Including figure files
\newcommand{\me}{\mathrm{e}}

\title[Eccentricity - Multiplicity Correlation]{Orbital Eccentricity - Multiplicity Correlation for Planetary Systems and Comparison to the Solar System}

\author[N. Bach-Møller, U.G.Jørgensen]{
Nanna Bach-Møller,$^{1}$\thanks{E-mail: nanna@bachmoeller.dk}
Uffe G. Jørgensen,$^{1}$
\\
$^{1}$Centre for Star and Planet Formation, Niels Bohr Institute, University of Copenhagen, Østervoldgade 5, 1350 Copenhagen, Denmark\\
}

\date{Accepted 2020 October 19. Received 2020 October 14; in original form 2020 March 18.}

\pubyear{2020}

\begin{document}
\label{firstpage}
\pagerange{\pageref{firstpage}--\pageref{lastpage}}
\maketitle

\begin{abstract}
The orbit eccentricities of the Solar System planets are unusually low compared to the average of known exoplanetary systems. A power law correlation has previously been found between the multiplicity of a planetary system and the orbital eccentricities of its components, for systems with multiplicities above two.
In this study we investigate the correlation for an expanded data sample, by focusing on planetary systems as units (unlike previous studies that have focused on individual planets). 
Our full data sample contains 1171 exoplanets, in 895 systems, and the correlation between eccentricity and multiplicity is found to follow a clear power law for all multiplicities above one. We discuss the correlation for several individual subsamples, and find that all samples consistently follow the same basic trend regardless of e.g. planet types and detection methods. We find that the eccentricities of the Solar System fit the general trend and suggest that the Solar System might not show uncommonly low eccentricities (as often speculated) but rather uncommonly many planets compared to a "standard" planetary system. 
The only outlier from the power law correlation is, consistently in all the samples, the one-planet systems. It has previously been suggested that this may be due to additional unseen exoplanets in the observed one-planet systems. Based on this assumption and the power law correlation, we estimate that the probability of a system having 8 planets or more is of the order of 1\%, in good agreement with recent predictions from analyses based on independent arguments.

\end{abstract}


	%
	%-------------------------------------------------------------------
	
\section{Introduction}
Extrasolar planets reveal orbital eccentricities much higher than those found among the planets of the Solar System, a deviation that in the beginning was considered so strange that it even lead some people to doubt whether the radial velocity exoplanet measurements actually showed real planets. In the present study we will show that the eccentricity of the Solar System planets actually follow the same trend as all other known planetary systems, but belong to the tail of a continuous distribution.
When searching for extraterrestrial life we often focus on Earth-like planets and Solar System-like systems, and so low eccentricities are included in our search criteria. But exactly how the habitability of a planet might be affected by the eccentricity of its orbit is yet unknown. A planet on a high-eccentricity orbit can undergo drastic seasonal changes in surface temperature due to the difference in stellar radiation from perihelion to aphelion. These seasonal changes could lead to periods of time without liquid water on the surface, which would greatly limit the habitability of the planet \citep{bolmont2016habitability}. However, a series of studies (reviewed in \cite{2019arXiv191104441K}) have found that often the atmosphere and oceans of a planet can act like a buffer to the temperature variations, in which case the surface climate will be determined by the average stellar radiation rather than the seasonal extremes. In other cases large seasonal variability was found to expand the habitable zone for the planet, by allowing water to remain liquid at larger semi major axes \citep{linsenmeier2015climate}. Since it is still uncertain how orbit eccentricities affect the habitability of a planet, it is critical for us to study and understand the eccentricities in the existing exoplanet sample and how they might deviate from those in the Solar System.
	
From previous investigations \citep{2007DDA....38.1501C,2008ApJ...686..621F,2008ApJ...686..603J,2008ASPC..398..295J, 2019A&A...629L...7C}, planet-planet interaction has been suggested as the dominating mechanism determining orbital eccentricities of planets, either through dynamical relaxation or planet-planet scattering. The dynamical interactions of planetary systems is reviewed in \cite{2014prpl.conf..787D}. As a conclusion of this, a correlation between orbital eccentricity and multiplicity (number of planets) is predicted. This prediction has been tested empirically by \citet{2015PNAS..112...20L} based on 403 exoplanets detected by the radial velocity method (RV) and listed in \textit{exoplanets.org}. A strong anti-correlation between eccentricity (e) and multiplicity (M) was found, and for multiplicities above two the correlation could be described by a power law: $e(M)\approx 0.584\cdot M^{-1.20}$. The eccentricity-multiplicity correlation has later been investigated by \cite{2017A&A...605L...4Z}, who found a similar correlation for multiplicities above one based on 258 selected RV and transit planets from NASA Exoplanet Archive. Both of the previous investigations have based their analyses on individual planets rather than treating the systems as units.
	
The main motivation for this article is to further the investigations by \citet{2015PNAS..112...20L} and \cite{2017A&A...605L...4Z} using the expanded planet sample known to date, comparing search methods, population groups, and databases, and aiming to set the results in perspective to our own Solar System and habitability. Our planet sample contains planets found by several detection methods including RV, transiting planet (transit), microlensing (ML) and others. By including all planets, regardless of detection method, we will be able to comment on whether there is an observational bias related to the specific methods, and the large dataset available today makes it possible to exclude more planets that might potentially introduce unwanted bias into the correlation. Unlike the previous investigations we will treat each system as a unit by conducting the analysis based on the average orbital eccentricities in the systems rather than the eccentricity of each individual planet. This is done since both the multiplicity and potential planet-planet interactions are properties of the planetary system as a whole rather than the individual planets. \\
From the resulting eccentricity-multiplicity correlation an estimate of the mean multiplicity of a planetary system can be obtained in addition to a probability distribution of the multiplicity of planetary systems. From this we wish to set our Solar System in perspective against a "standard" planetary system.
We envision that planetesimals are formed in relatively circular orbits, then gravitationally scatter one another into higher eccentricity, before they over longer timescales collide to build up solid planets or the planetary cores of giants. After the evaporation of the gas disk, planet-planet interaction would be the dominating mechanism determining the final eccentricities, in such a way that the more planets there end up being in the system the more circular the orbits become. This is a logic scenario to provide an image of the physical process behind the correlation we investigate in the present paper, but we stress that this is only an image that helps us (and hopefully the reader, too) to imagine the process. Our study is empirical, and hence have no apriori assumption about which exact mechanisms cause the correlation. In order to further the development of the theoretical understanding, we take advantage of the large sample now available to also analyze whether different populations of exoplanets show different correlations.\\

A major concern when investigating extrasolar planets is that we are highly constrained by limitations in our detection methods. When using RV the detection probability of a planet is biased towards large masses, and when using transit it is biased towards ultra short periods. That leaves a large parameter space where planets go mainly undetected, and thereby bias conclusions about standard planetary systems drawn from the limited sample. Today the two most abundant detection methods (RV and transit) basically have shown us that exoplanetary systems very different from our own Solar System are abundant. Direct observational estimates of how abundant exoplanetary systems resembling our own Solar System are, may most likely come from future extensive microlensing surveys from space (perhaps from a dedicated microlensing satellite \citep{Bennett_2002} or from WFIRST \citep{penny2019predictions}) or from the ground (perhaps from GravityCam-like instruments \citep{mackay2018gravitycam}), and they will give us the full set of orbital parameters of solar-system-like exoplanets \citep{gaudi2012microlensing,2018AJ....155...40R}, as opposed to today where orbital eccentricity has been obtained for only one microlensing exoplanet \citep{gaudi2008discovery}. Until then it can be useful to look at indirect evidences for what a standard exoplanetary system looks like. A motivation for this article is to go beyond the data sample by finding a general theory for all systems (including those with planets yet undetected), and from this estimate the characteristics of standard planetary systems. This may give us some insight into the standard formation mechanism of planetary systems and how they develop into the most common configurations of planets, give hints about what to look for and thereby which instruments to develop, and maybe contribute to give us a more realistic view on how abundant truly Earth-like exoplanets might be. One such indirect method is the study of the eccentricity distribution among known exoplanets, as presented here.\\
	
In Sect.\, \ref{sec:Data} the dataset is discussed.
In Sect.\, \ref{sec:e(M)} the correlation between eccentricity and multiplicity is examined, both for the full data samples from two different databases, for subsamples sorted for detection methods and for population groups, and for a high-eccentricity subsample in which we attempt to exclude most systems containing undiscovered planets. Based on the correlation a power law is found. In Sect.\, \ref{sec:meanM} some of the potential implications of the power law correlation are explored. A probability distribution of the multiplicity is found, and from this a mean multiplicity of planetary systems is estimated. In Sect.\, \ref{sec:Dis} the results and theories are discussed. Finally in Sect.\, \ref{sec:Con} the conclusions are summarized.

	%--------------------------------------------------------------------
\section{The Dataset} \label{sec:Data}	
Our data from \textit{exoplanet.eu} were retrieved in August 2019. All confirmed planets regardless of detection method are included. We are aware that \textit{exoplanet.eu}, like most other databases, might be subject to errors in their data listing. For the sake of this study we mostly try not to question the validity of the data found on the website. Planets without listed eccentricities or where the eccentricity is falsely listed as zero (i.e. without listed uncertainties) are excluded from the sample. Of the 4103 planets listed on \textit{exoplanet.eu} a total of 1171 planets remain in the sample, 2932 are excluded due to unknown eccentricities and 60 of these have eccentricities listed as zero with unknown uncertainties. In Table \ref{tab:M}\footnote{All planets and systems with a multiplicity of X will henceforth be referred to as MX-planets or MX-systems} the number of planets sorted by multiplicity can be seen for each of the included detection methods.

Because no multiplicities are listed on \textit{exoplanet.eu} each planet has been given a multiplicity based on the number of confirmed planets orbiting the same star listed in the database. Since some of the systems might contain yet undiscovered planets the known companions in these systems will initially be sorted into the wrong multiplicity bins, and the actual distribution might differ from Table \ref{tab:M}. Due to the small number of systems with high multiplicities, all systems with more than 5 known planets have been combined in one bin. The multiplicity of this bin is calculated as the mean multiplicity of the included systems.
Note that the number of planets in each bin is not necessarily a multiple of the multipliciticy. This is caused by the fact that not all planets in each system are included, mainly because their eccentricities are unknown. Our dataset is three to four times larger than any of the previous analyses (1171 in this study, compared to 403 in \cite{2015PNAS..112...20L} and 258 in \cite{2017A&A...605L...4Z}). We have not accounted for the uncertainties listed for each of the eccentricities in the database in this analysis, which will be discussed further in Sec. \ref{sec:Dis}.

\begin{table}
	\caption[]{Planets included in data samples. Retrieved from $exoplanet.eu$. Planets are sorted for detection method. Rightmost column show the number of systems present in each multiplicity bin, whereas columns 2-5 show number of individual planets.}
	\label{tab:M}
	$$ 
	\begin{array}{p{0.2\linewidth} l l l l | l}
	\hline
	\noalign{\smallskip}
	Multiplicity & Total & RV & Transit & Other & Systems  \\
	\noalign{\smallskip}
	\hline
	\noalign{\smallskip}
	M1 & 667 & 408 & 234 & 23 & 667  \\
	M2 & 274 & 215 & 52 & 5 & 151 \\
	M3 & 121 & 65 & 50 & 6 & 45 \\
	M4 & 63 & 43 & 17 & 3 & 20 \\
	${\geq}$M5 & 46 & 34 & 10 & 2 & 12\\
	\noalign{\smallskip}
	\hline
	\noalign{\smallskip}
	Total & 1171 & 765 & 363 & 39 & 895\\ 
	\noalign{\smallskip}
	\hline
	\end{array}
	$$
\end{table}

%--------------------------------------------------------------------

\section{Eccentricity and multiplicity} \label{sec:e(M)}

\begin{figure}
	\centering
	\includegraphics[width=\columnwidth]{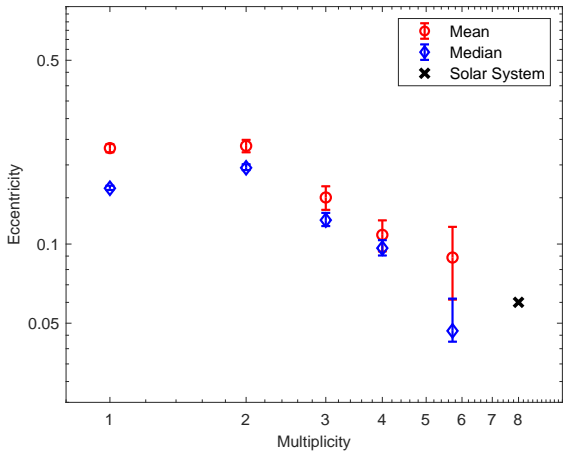}
	\caption{ Mean and median values of the eccentricity for each multiplicity. The mean eccentricity of the Solar System is plotted with a black $\times$. The multiplicity of the ${\geq}M5$ multiplicity-bin is plotted as $M=5.7$.}
	\label{e(M)}
\end{figure}

Each system is assigned an eccentricity found as the mean eccentricity of the planets in the system. This differs from previous studies, where the planets were not sorted into systems, and the authors looked at the eccentricities of the individual planets. The final results from the two methods do not differ greatly, but we find that sorting the planets into systems is more meaningful, since the effects we observe might be caused by planet-planet interactions within the systems and will change the system as a whole. 
These assigned system eccentricities are then used to calculate overall mean and median eccentricities within each multiplicity bin.
In Fig.\, \ref{e(M)} mean and median values of the system eccentricities are plotted for each of the multiplicity bins, together with our Solar System with a multiplicity of eight. 

The errors are calculated using the following methods: Mean; As the standard deviation of system means found by the Bootstrap method. Median; As the one-third and two-thirds quantiles from a Cumulative Distribution Function divided by $\sqrt{N-1}$, where $N$ is the number of systems in the multiplicity bin. Notice that the errors indicate the uncertainties of the mean and median eccentricities of each multiplicity bin, and not the spread of the eccentricities among the individual planets, which is significantly larger than the errors shown.
Fig.\, \ref{e(M)} suggests a trend of decreasing eccentricity for increasing multiplicity. As can be seen the Solar System too follows this trend indicating that our system does not deviate from the norm.
An exception for this trend, is the M1 systems. Whereas the other data points seems to approximately follow a power law (seemingly linear because of the logarithmic axes), the eccentricities for M1 deviate from the trend by being too low to follow the power law. This deviation will be discussed later.

\subsection{Planet populations} \label{sec:pop}

A potential uncertainty related to the study of an eccentricity-multiplicity correlation is the dependence of the correlation on factors such as planet mass and semi major axis. \citet{turrini2020normalized} and \citet{laskar2017amd} therefore looked at the correlation of multiplicity and angular momentum deficit (AMD), rather than multiplicity and eccentricity. The AMD does depend on the eccentricity, but also on the semi major axis and the mass of the planets, and \citet{turrini2020normalized} found an anticorrelation between the normalized angular momentum deficit (NAMD) and the multiplicity. \citet{turrini2020normalized} argues that the eccentricity-multiplicity correlation found by other studies is a reflection of the underlying NAMD-multiplicity correlation. The study of the NAMD-multiplicity is complicated by the fact that few planets have both their masses, eccentricity and semi-major axis well-known, and as such the dataset is smaller. The larger sample in our data set compared to previous data sets, allows us to study directly the correlation of eccentricity and multiplicity for a number of different subsamples, in order to test how the planet mass ($m_p$) and semi major axis (or period, $P$) might affect the eccentricity-multiplicity correlation.\\

To test the impact of mass and period, we have divided the systems into three different populations: 1) Systems containing a hot-Jupiter ($m_p > 0.1 M_J$ and $P < 100$  days). 2) Systems containing a cold-Jupiter ($m_p > 0.1 M_J$ and $P > 100$  days), and no hot-Jupiters. 3) Systems dominated by super-Earths ($m_p < 0.1 M_J$) with no giant planets. In order to increase the data sample, planets with no listed mass in the database, have been sorted based on their $mass\cdot sin(i)$ value, when this is known, and a total of 849 systems are sorted into the population categories. The distribution of systems in each population category can be seen in Table \ref{tab:pop}. It should be noted, that the observed planet sample does not represent the true planet population since some planet types are more easily observed than others, but the differences between the populations, as shown here, might still give us an insight into the uncertainties of the eccentricity-multiplicity correlation. Research in the actual occurrence rate of different planet types is reviewed in e.g. \cite{winn2015occurrence}.

\begin{figure}
	\centering
	\includegraphics[width=\columnwidth]{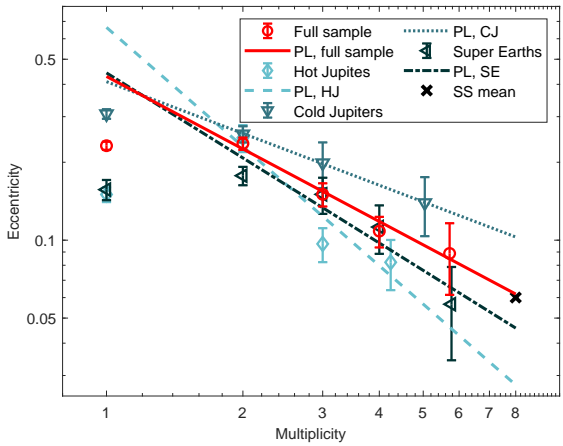}
	\caption{ Mean values of eccentricities for each multiplicity for four subsamples. Full red line: The full sample from \textit{exoplanet.eu} identical to the mean values from Fig.\, \ref{e(M)}. Dashed: Subsample of systems containing a hot-Jupiter (HJ). Dotted: Subsample of systems containing a cold-Jupiter (CJ). Dot-dashed: Subsample of systems only containing smaller planets. Mean value of the Solar System (SS) is plotted in black. Power laws (PL) have been fitted to all four samples for multiplicities above one; this is discussed in Sect.\, \ref{sec:pop}.}
	\label{fig:pop}
\end{figure}

Table \ref{tab:pop} shows that different multiplicities are dominated by different populations of planets, such that most of the M1 systems are giant-planet systems, whereas the larger multiplicity systems are dominated by super-Earths. A priori one could expect that since the cold-Jupiters dominate the M1 systems, we could seek the explanation for the deviation from the power law followed by the $M>1$ systems in the cold-Jupiter population. However, we find that this is not the case, when we look at the mean eccentricities plotted as a function of multiplicity in Fig.\,\ref{fig:pop}.

\begin{table}
	\caption[]{Distribution of the systems in Table. \ref{tab:M} where in addition to the eccentricity, also the mass or $m sin(i)$ is known, such that they can be divided into the groups: hot-Jupiters (HJ), cold-Jupiters (CJ), super-Earths (SE), and plotted in Fig. \ref{fig:pop}. Last column shows the number of systems (Number). A total of 849 systems are included.}
	\label{tab:pop}
	$$ 
	\begin{array}{p {0.2\linewidth} l l l | l }
	\hline
	\noalign{\smallskip}
	Multiplicity & HJ \hspace{0.08\linewidth}& CJ \hspace{0.08\linewidth}& SE \hspace{0.1\linewidth}& Number \\
	\noalign{\smallskip}
	\hline
	\noalign{\smallskip}
	M1 & 39.2\% & 51.3\% & 9.4\% & 637 \\
	M2 & 22.9\% & 57.6\% & 19.4\% & 144 \\
	M3 & 25.6\% & 20.5\% & 53.8\% & 39 \\
	M4 & 21.1\% & 26.3\% & 52.6\% & 19 \\
	${\geq}$M5 & 10.0\% & 30.0\% &  60.0\% & 10 \\
	\hline
	\end{array}
	$$
\end{table}

Fig.\,\ref{fig:pop} shows the mean eccentricities plotted for the full sample (equivalent to the mean values from Fig.\, \ref{e(M)}) together with the three different populations introduced above. A power law has been fitted to all samples for multiplicities above one, not including the Solar System, i.e. $1<M<8$. The power law has been fitted to the overall mean eccentricities for all systems in each multiplicity bin, corresponding to the data points seen in the figure. Due to the small sample of Jupiter-systems with four or more planets, the $M4$ and ${\geq}M5$ bins have been combined for the hot-Jupiter and cold-Jupiter systems. The multiplicity for these bins are the mean multiplicities among the systems combined in the bins.
The main conclusion from Fig.\,\ref{fig:pop} is that all three populations follow similar power law trends to the one for the full sample (although of course with larger scatter of the individual points due to the smaller data sample). We notice that the cold-Jupiter population is not the cause of the low eccentricities of the M1 systems, but on the contrary displays the highest eccentricities of the M1 systems among all populations.

\subsection{The undiscovered planets in the systems}

\begin{figure}
		\centering
		\includegraphics[width=\hsize]{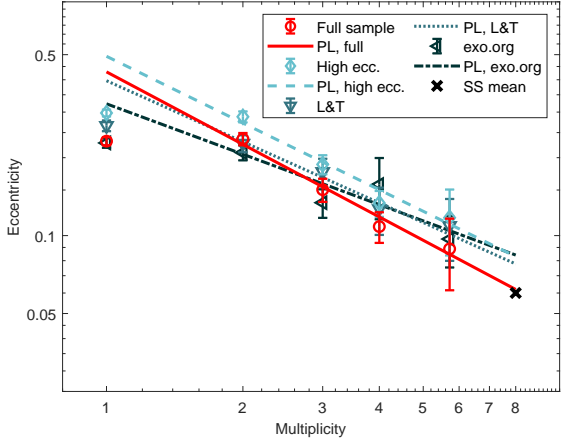}
		\caption{Mean values of eccentricities for each multiplicity for four subsamples. Full red line: The full sample from \textit{exoplanet.eu} identical to the mean values from Fig.\, \ref{e(M)}. Dashed: High-eccentricity subsample consisting of 75\% systems with highest eccentricities. Dotted: Subsample of RV planets detected before 2014 equivalent to the sample used by \citet{2015PNAS..112...20L}. Dot-dashed: Full sample from \textit{exoplanets.org}. Mean value of the Solar System (SS) is plotted in black. Power laws (PL) have been fitted to all samples for multiplicities above one; this will be discussed in Sect.\, \ref{sec:meanM}.}
		\label{e(M)full75LT}
	\end{figure}

To get further understanding of the uncertainties of the power law correlation, Fig.\,\ref{e(M)full75LT} shows the mean eccentricities plotted as a function of multiplicities for three additional subsamples: Beside the full system sample from \textit{exoplanet.eu}, are shown a high-eccentricity subsample consisting of only the 75\% systems with highest eccentricities, a subsample consisting of RV planets listed on \textit{exoplanets.org} before 2014 (L\&T) equivalent to the sample used by \citet{2015PNAS..112...20L}, and a full sample of the 704 planets with known eccentricities from the database \textit{exoplanets.org}. Power laws have been fitted to all samples for multiplicities above one. \\

The high-eccentricity subsample has been created to exclude systems containing undiscovered planets.
According to the trend visible in Fig.\, \ref{e(M)} larger systems have lower eccentricities, and systems with additional, undiscovered, planets should therefore have eccentricities below what is appropriate for their given multiplicity. We might therefore expect, that the systems showing the lowest orbital eccentricities, could have extra undiscovered planets. 
Removing these systems from the fit does change the relation a bit (obviously shifting the line to somewhat higher eccentricities), but do keep the same trend of a fine linear fit to the systems with $M>1$ and a substantially lower average eccentricity for the M1 systems than expected from the power law.

Since both of the dominating detection methods (the radial velocity method and the transit method) depend on the size of the planets, smaller planets are more difficult to detect, and only few planets with a size comparable to Mercury or Mars have been found. Mars and Mercury represent one fourth of the (known) planets in the Solar System, and following this line of argument a first attempt of a qualified guess on a typical number of undetected planets could be, that a minimum of 25\% of the planets in exoplanet systems remain undiscovered. By removing the 25\% systems with the lowest eccentricities in each multiplicity-bin we hope to lower the bias in the correlations by "contamination" due to systems with unknown planets.
No systems are removed from the M8 bin, since this only consist of the Solar System. We see from Fig.\, \ref{e(M)full75LT} that the high-multiplicity systems are less affected than the low-multiplicity systems when removing the 25\% lowest eccentricity systems, indicating that high-multiplicity systems could be more completely surveyed.\\

The L\&T subsample has been plotted to compare the power law correlation found in this study with one found using a data sample similar to the one used in the original study by \citet{2015PNAS..112...20L}. Notice that whereas the mean eccentricities for the full, high-eccentricity, and \textit{exoplanets.org} subsamples are found as the mean of the system eccentricities for each multiplicity, the mean eccentricities of the L\&T subsample are found as the mean of all planets in each multiplicity-bin (to stay consistent with the analysis methods used by \citet{2015PNAS..112...20L} as explained previously). \\

In order to further constrain potential uncertainties related to our data, we repeated the entire analysis using data from the database \textit{exoplanets.org}. It should be remembered that our main database, \textit{exoplanet.eu}, is more complete and up to date than \textit{exoplanets.org}, but that the planets listed on \textit{exoplanets.org} have undergone a more strict selection process in regard to peer-review (\cite{2014PASP..126..827H,schneider2011defining}, and personal communication with Jason Wright and Françoise Roques). Although the two databases therefore will not contain the exact same data sample, comparison of the results based on both databases gives more clear impression of the uncertainties. \\

Fig.\, \ref{e(M)full75LT} shows that all the subsamples, display the same general tendency of a power law correlation between eccentricity and multiplicity for $M>1$ as the full sample, and a lower eccentricity of the M1 systems not following the power law trend of the higher multiplicity systems. The slopes, however, vary for the different samples.

\begin{figure}
	\includegraphics[width=\hsize]{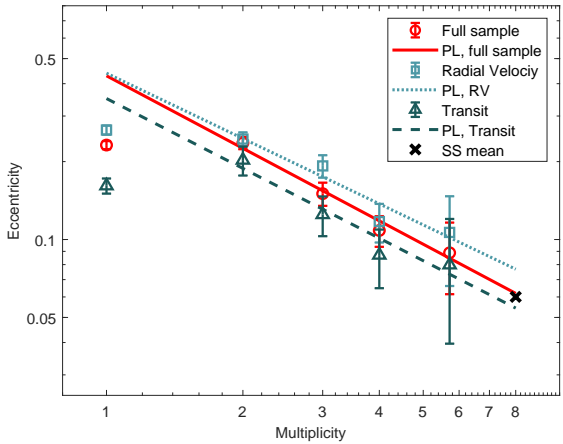}
	\caption{Mean values of eccentricities for each multiplicity for three subsamples. Full red line: The full sample. Dashed: Subsample consisting of planets discovered by the transit method. Dotted: Subsample consisting of planets discovered by RV. Mean value of the Solar System (SS) is plotted in black. Power laws (PL) have been fitted to all samples for multiplicities above one.}
	\label{e(M)met}
\end{figure}

\subsection{Detection methods}
Whereas the L\&T subsample consists only of RV planets our sample contains planets found by all detection methods. To test how this difference might affect the eccentricity-multiplicity correlation, and to better understand whether the behaviour of the correlation could be dominated by a bias effect related to the detection method, a plot for the transit and RV subsamples together with the full sample can be seen in Fig.\, \ref{e(M)met}. It should be noted that the eccentricities listed for planets discovered with the transit method are often determined from followup RV observations, so the two populations are not completely separated.
Fig.\, \ref{e(M)met} shows that both the transit and the RV subsamples have eccentricity-multiplicity correlations similar to that of the full sample, and the trend of the M1 systems falling below the $M>1$ relation is identical. 

We also see that the transit systems show lower eccentricities at all multiplicities compared to the RV systems. This bias, that transit planets generally have lower eccentricities, is in correspondence with a study by \citet{van2015eccentricity} who found high-multiplicity Kepler planets to generally have lower eccentricities than the RV planet sample. This tendency might be caused by the bias, that there are more low-mass planets in the transit subsample than in the RV sample, and that lower mass planets are more easily circularised by planet-planet interaction \citep{kane2012exoplanet}. We see a hint of the same tendency in Fig.\,\ref{fig:pop} where the super-Earth subsample shows lower eccentricities than the full sample, and the important conclusion is that independent of the shift and its potential explanation in an observational bias, the same tendencies discussed above applies to both of the subsamples.

It is also possible that planet-planet scattering could cause a spread in the orbital inclinations \citep{2007DDA....38.1501C} in addition to lowering the multiplicity of the system. The spread in inclination could lead to a higher number of undiscovered planets in the transit systems and thereby a higher number systems with eccentricities too low to fit their assigned multiplicity. This trend would be strongest for low-multiplicity systems, as seen in Fig.\, \ref{e(M)met}, if these are formed due to severe planet-planet scattering. 
It can be seen from the errorbars given in Fig.\, \ref{e(M)met} that the listed eccentricities of the transit planets have a greater variation than the RV planets, possibly caused by a larger uncertainty in their determination \citep{kane2012exoplanet,van2015eccentricity}.

\subsection{Kolmogorov-Smirnov test}
To statistically test the correlation between multiplicity and eccentricity, a two-sample Kolmogorov-Smirnov is conducted on the full system sample. The test compares the multiplicity-bins one and one to test the difference in the eccentricity distributions of the systems. The test results can be seen in Table \ref{tab:KS}. Notice that the distribution of eccentricities for the individual \textit{planets} is used for the Solar System, whereas the distributions of the \textit{systems} are used for the rest.\\
It can be seen that the eccentricities of most of the multiplicity-combinations show significant differences, on a 5\% significance level. This indicates that the difference in eccentricity for systems of different multiplicity is caused by a connection between the two factors and not by coincidence. The higher p-values seen for high-multiplicity combinations might be caused by the small number of systems in these multiplicity-bins.
Altogether the statistical test supports, that there is a correlation between multiplicity and eccentricity. \\

\begin{table}
	\caption[]{Test result for Kolmogorov-Smirnov test.}
	\label{tab:KS}
	$$ 
	\begin{array}{p{0.15\linewidth} l l l l l l}
	\hline
	\noalign{\smallskip}
	& M1 & M2 & M3 & M4 & {\geq}M5 & M8 (SS)  \\
	\noalign{\smallskip}
	\hline
	\noalign{\smallskip}
	M1 & 1 &  &  &  &  &  \\
	M2 & <0.01 & 1 &  & & & \\ 
	M3 & 0.04 & 0.01 & 1 & & & \\ 
	M4 & 0.01 & <0.01 & 0.15 & 1 & & \\
	$\geq$M5 & 0.04 & <0.01 & 0.06 & 0.31 & 1 & \\
	M8 (SS) & 0.01 & <0.01 & 0.05 & 0.38 & 0.65 & 1 \\
	\noalign{\smallskip}
	\hline
	\end{array}
	$$ 
\end{table}

\subsection{Quantification of the multiplicity-eccentricity correlation}
In the standard core-accretion model for the formation of planetary systems, the dust component of the disk relatively quickly clumps together (via simple condensation or even faster via streaming instability) to form many objects of planetesimal sizes \citep{johansen2017forming}. Over a longer timescale the planetesimals then excite one another’s orbits by gravitational interaction, leading to collisions and hence growth to planet size. After the dissipation of the protoplanetary disk the orbits of the planets are largely determined by planet-planet interactions, indicating a correlation between the orbital eccentricity and the number of interactions and hence planets. The numerical simulations by \cite{2007DDA....38.1501C} and \cite{2008ASPC..398..295J} confirms that this expectation is correct, by showing that the final architecture of a system is almost independent of the assumed starting conditions of planetesimals, and suggesting that planet-planet interaction is the dominating mechanism for changing the average orbital eccentricity. The simulations do not in themselves predict a specific analytical correspondence between eccentricity and multiplicity, which, however, can be done by fitting the corresponding observational data. In Fig.\, \ref{e(M)full75LT} it was indicated that the high-multiplicity systems seemed to have fewer undiscovered planets, and in both Fig.\, \ref{fig:pop}, \ref{e(M)full75LT} and \ref{e(M)met} we quantified the relation by fitting the mean eccentricities for $M>1$ to a power law. Our best fit to the full set of data (as shown in red in the figures) can be expressed as:
\begin{equation}
e(M)=0.429\cdot M^{-0.93}
\label{eq:e(M)}
\end{equation}
where $e$ is the eccentricity and $M$ is the multiplicity. Fig.\, \ref{fig:pop}-\ref{e(M)full75LT} and \ref{e(M)met} further demonstrates that this fit also agrees with the Solar System despite the fact that the $M=8$ was not included in the fit. This adds extra confidence in believing that the quantification is universal, and two fits, with and without the Solar System, showed the following correlation coefficient; $R^2=0.98$ for $M=[2;7]$ and $R^2=0.99$ for $M=[2;8]$. \\

Since the physical cause behind the relation is thought to be planet-planet gravitational interaction, one should expect the decreasing tendency to range all the way from M1 systems to a maximum number of planets, $M_{\rm max}$, for which the systems can still remain stable, \citep{2001MNRAS.325..221P,2008ApJ...686..603J},
with the M1 systems having the largest average eccentricity. Observationally, the M1 planets, obviously, do not show the high eccentricity expected from the correlation, and therefore the observed M1 systems must be affected differently from the multi-planet populations. In the following section, Sect.\, \ref{sec:meanM}, we will elaborate on one potential explanation for the deviation of the M1 systems from the trend, namely the idea that the low M1 eccentricity is caused by a combination of mechanisms other than the general planet-planet interaction, lowering the eccentricities, plus an observational bias. When correcting for these two effects, the remaining M1 systems are made to follow the same trend as the rest of the systems, and potential implications for the trend are explored. 

An alternative explanation for the discrepancy between the M1 and multi-planet systems could be that they are dominated by different planet populations. To analyze if any specific population dominates the lowering of the M1 eccentricities, we investigated, in Sect.\, \ref{sec:pop}, whether the population of large planets (which observationally dominates the M1 and M2 systems) and the population of smaller planets (that have a more dominating role in the higher multiplicity systems), show different observational trends. We concluded that all of the populations follow the same general trend between eccentricity and multiplicity, indicating that the same general mechanism is responsible for all the observed populations of exoplanets from M1 to M8 (and is likely to be planet-planet interaction with some correction for the M1 systems).

In all cases, it is obvious from Fig.\, \ref{e(M)}-\ref{e(M)met} that the observed M1 systems do not follow the trend expressed in Eq.\, \ref{eq:e(M)}. If a reasonable transformation from the observed abundance of M1 systems to intrinsic M1 system abundances can be obtained, it will be possible from Eq.\,\ref{eq:e(M)} to give an estimate of the true probability distribution of multiplicities among the observed systems.

%--------------------------------------------------------------------
\section{Perspective and implications: Conversion of observed multiplicity distribution to actual distribution} \label{sec:meanM}

Fig.\, \ref{e(M)}-\ref{e(M)met} demonstrates that the observed average eccentricity of one-planet systems (M1) falls below the relation for multi-planet systems. The main assumption in this further analysis is that the M1 systems intrinsically follow the same eccentricity correlation as the other multiplicities. This assumption is supported by a series of studies by \cite{he2019architectures,he2020architectures}, who recreated the multiplicity distribution of the Kepler observations, by forward-modelling multi-planet systems at the AMD-stability limit (introduced in \citet{laskar2017amd,petit2017amd}). \cite{he2020architectures}, found that all multiplicites from one to ten followed the same eccentricity-multiplicity power law correlation, with the intrinsic M1 systems having higher eccentricites than the multi-planet systems, and they found that most observed M1 systems contain yet undiscovered planets. In this section will will try to identify these systems with undiscovered planets, and redistribute them to the multiplicity bin appropriate to their multiplicities.\\

We will first investigate whether some of the low eccentricity M1 planets can have got their low eccentricity due to other mechanisms than the general planet-planet interaction assumed to be responsible for Eq.\,\ref{eq:e(M)}. \\
Exoplanets in ultra small orbits are often tidally locked to the host star, which could lead to circularisation of the planetary orbit \citep{2008IAUS..249..187J}. By looking at the eccentricity damping timescale \citep{2014ARA&A..52..171O}, the eccentricity damping from these planet-star interactions can be approximated by:
\begin{equation}
\dot{e} \propto \frac{m_*}{m_p}\frac{1}{a^5}
\label{eq:edamp}
\end{equation}
where $\dot{e}$ is the change in eccentricity, $a$ is the semi major axis of the planet, and $m_p$ and $m_*$ are the masses of the planet and the star respectively.

In order to distinguish systems that have low eccentricities due to planet-star interactions from those that may have low eccentricities for other reasons, all planets for which the value from Eq.\, \ref{eq:edamp} exceeds a certain threshold are excluded. The threshold was chosen to $6.77\times10^5$, and 191 M1 planets, and 100 planets among the other multiplicities, were excluded on this basis. These planets will be excluded in the following probability analysis, but were not excluded in the making of Eq. \ref{eq:e(M)} (which would have very small effect as described below). The chosen threshold is the value of Mercury, and even though Mercury is far from being circularised (it holds the highest eccentricity in the Solar System), it is "almost" tidally locked (in a 2/3 orbital/rotational resonance), and is the planet in the Solar System that has the highest potential for tidal circularisation. In an analysis of hot-Jupiters with known obliquities, \cite{Hjortphd} was able to divide the planets into two distinct groups, with 15\% of the planets having extremely low obliquity (and hence low eccentricity) and 85\% having a continuous obliquity distribution. \cite{Hjortphd} ascribed the former group to planet migration in the disk and the latter to migration due to planet-planet interaction (scattering). It is therefore likely that also a fraction of the M1 systems will have much lower eccentricities than expected from Eq.\, \ref{eq:e(M)} due to disk-migration.\\ 

Next, we pursue the idea that some of the remaining systems may contain yet undiscovered planets, and that these systems will lower the mean eccentricity of their multiplicity bins, since systems with more planets are expected to have lower eccentricities. Those of the observed systems that have had their eccentricity determined by planet-planet interactions (as opposed to the systems excluded above due to  a potential star-planet circularisation) are to first approximation expected to follow the planet-planet eccentricity relation expressed in Eq.\,\ref{eq:e(M)}. We align the mean eccentricities of the multiplicity bins with the power law correlation, by moving the lowest eccentricity systems of the multiplicity bins to an M corresponding to their observed eccentricity (i.e. assuming undiscovered planets in those systems). During this exercise it was found, that the best alignment occurred when 55\% of the M1 systems and all of the $M>1$ systems were assumed \textit{not} to contain undiscovered planets, and the rest had new multiplicities estimated based on their eccentricity.

Of the M1 systems 50 (i.e. roughly 10\% of the $667-191=476$ M1 systems that remained after the exclusion of planets that might have experienced planet-star circularisation) have such low eccentricities that they should be moved to multiplicities that might exceed M$_{\rm max}$ (in some cases more than 50 planets). It was therefore assumed that these systems might not contain undiscovered planets, but that other physical mechanisms were responsible for circularizing these 10\%.
 For the proceeding estimates, have in mind that the effect of keeping these 50 planets would be to slightly increase the estimated abundance of M1 systems and decrease correspondingly the abundance of high multiplicity systems like our own solar system.
Non-planet-planet interacting mechanisms that could be responsible for circularization of a fraction of this amount of M1 systems could include migration of a single large planet to small orbit while substantial amount of the protoplanetary disk was still in place (\cite{Hjortphd}). 

For those of the remaining group of ($667-191-50 = 426$) M1 systems with eccentricities that potentially could be attributed to yet undiscovered planets, we attempted a redistribution of the systems by artificially counting them as belonging to higher values of M. The new multiplicity, $M_{\rm new}$, was determined from the eccentricity of the planet using Eq.\, \ref{eq:e(M)}. A total of 164 M1 systems were redistributed and the new multiplicity distribution can be seen in Table. \ref{tab:newM}. 

\begin{table*}
	\caption[]{Redistribution of systems. Left; the observed multiplicity distribution of systems from $exoplanet.eu$. Right; the multiplicity distribution of systems after the M1 systems have been redistributed according to their eccentricities as described in the text. The rightmost column indicates the probability of a system having a given multiplicity according to Eq.\, \ref{eq:prob2}} 
	\label{tab:newM}
	$$ 
	\begin{array}{l l l l l l}
	\hline
	\noalign{\smallskip}
	Multiplicity & \multicolumn{2}{l}{Observed\: distribution} & \multicolumn{3}{l}{Redistribution}  \\
	\noalign{\smallskip}
	\hline
	\noalign{\smallskip}
	   & Number \: of \: Systems & Percentage & Number \: of \: Systems & Percentage & Probability \\   
	M1 & 667 & 75\%   & 262 & 41\% & 41\% \\
	M2 & 151 & 17\%   & 149 & 24\% & 24\% \\ 
	M3 & 45  & 5\%    & 90  & 14\% & 14\% \\
	M4 & 20  & 2\%    & 53  & 8\%  & 8\% \\ 
	M5 & 4   & $<1$\% & 25  & 4\%  & 5\% \\
	M6 & 6   & $<1$\% & 21  & 3\%  & 3\% \\
	M7 & 1   & $<1$\% & 12  & 2\%  & 2\% \\
	M8 & 1   & $<1$\% & 7   & 1\%  & 1\% \\
	M9 &     &        & 5   & $<1$\%  & $<1$\% \\ 
	M10 &    &        & 9  & 2\%  & $<1$\% \\
	\noalign{\smallskip}
	\hline
	\noalign{\smallskip}
	Total & 895 & & 633 & &  \\
	\noalign{\smallskip}
	\hline
	\end{array}
	$$ 
\end{table*}

In addition to the number of systems within each multiplicity bin, Table \ref{tab:newM} also shows the percentage- and probability distributions for the redistributed planets. The probability distribution is found by fitting an exponential fit to the percentage distribution as shows in Fig. \ref{fig:pop} and explained later.

\begin{figure}
	\centering
	\includegraphics[width=\hsize]{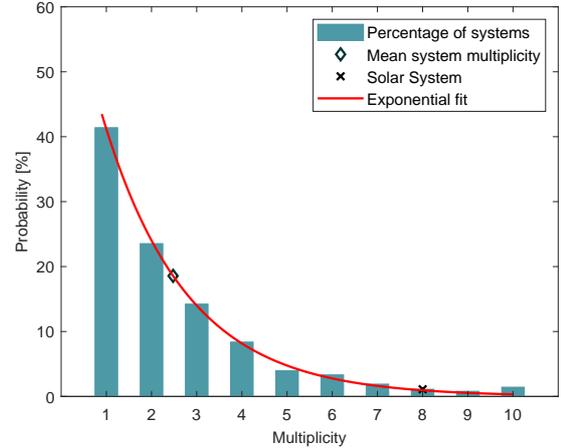}
	\caption{Percentage of systems with given multiplicity, corresponding to values from Table \ref{tab:newM}. Probability function found as exponential fit. Mean multiplicity estimated to $\sim 2.5$.}
	\label{prob}
\end{figure}

The redistribution of the M1 systems has been made such that the mean eccentricity of the remaining 262 systems falls on the same relation as the rest of the multiplicity systems described by Eq.\, \ref{eq:e(M)}. For the sake of this experiment, we assume that these remaining M1 systems would be the intrinsic M1 population among the observed systems, with no additional undiscovered planets and whose eccentricity is determined by the same planet-planet interactions as the multi-planet systems. In this sense one can think of the relation given by Eq.\,\ref{eq:e(M)} applied to all the systems from M1 to M$_{\rm max}$ as giving a minimum abundance of M1 systems and corresponding maximum abundance of high multiplicity systems. We stress this fact because it for many might seem intuitively (for example based on the antropic principle) surprising that our solar system belongs to such a relatively rare type of planetary systems as predicted from Eq.\,\ref{eq:e(M)} and shown in Fig.\,\ref{prob}; without the redistribution suggested above, the Solar System would be predicted to be of an even more rare type of planetary system.  

We therefore suggest that the $M_{\rm new}$ distribution in Table 3 is a reasonable first qualified guess of the relative distribution of the number of planets in planetary systems, whose average eccentricity distribution is determined by planet-planet interactions. This probability distribution is shown in Fig.\, \ref{prob} and has been fitted to an exponential function described as: 
\begin{equation}
P(M)=0.72\cdot \me^{-0.54M}
\label{eq:prob2}
\end{equation}
Where $P(M)$ indicates the probability of a system having $M$ planets. This relation has been found by normalizing the exponential fit seen in Fig.\, \ref{prob}, such that $\sum^{10}_{M=1}{P(M)}=1$.

The average number of planets in planetary systems, according to the distribution in Table. \ref{tab:newM}, is  $\langle M \rangle = 2.48$, and is marked by a diamond in the figure.
Based on the discrete probability distribution in Eq.\, \ref{eq:prob2} the probability of a system having eight planets is $P(8) \approx 1 \%$, indicating that systems the size of the Solar System are rare but not exceptionally so. In this interpretation the Solar System is in the tail of a distribution of multiplicity, and corresponding orbital eccentricities, near the maximum possible from stability considerations. We have in the above summation assumed that the maximum cut off is near to $M_{\rm max}$ = 10 planets, but we note that the exact value of $M_{\rm max}$ is unimportant for the conclusion, since the integral of Eq.\,\ref{eq:prob2} from 8 to infinity is very small. Remark also that the number 1\% refers to the fraction of the systems that have had their eccentricities determined by planet-planet interaction, or a similar process that is responsible for Eq.\,\ref{eq:e(M)} and Eq.\, \ref{eq:prob2}. If one is counting also the M1 planets that were excluded in deriving Eq.\ref{eq:prob2}, then the probability of finding 8 planets would be slightly lower. It should be noted that all results in this analysis rely on the assumption that the power law in Eq. \ref{eq:e(M)} describes the true intrinsic correlation between eccentricity and multiplicity. The redistribution was based on a correlation fitted to the observed multi-planet systems. The fact that some of these multi-planet systems might host yet undiscovered planets could therefore pose a uncertainty to the analysis. However, theoretical studies have found that the observed M1 population is the only one that differ greatly from the theoretical predictions \citep{johansen2012can}. As mentioned previously, some have suggested that this is caused by the fact that the observed M1 systems are especially prone to containing undiscovered planets \citep{he2020architectures}. As such, the analysis should not be affected greatly by undiscovered planets in the multi-planet systems. As mentioned previously planet-star interaction has not been taken into account when making Eq. \ref{eq:e(M)}. If the planets from the multi-planet systems that might have experienced planet-star interaction had been excluded in Eq. \ref{eq:e(M)}, the mean multiplicity would have been 2.6 rather than 2.5 planets per system. \\
It is encouraging to note that \cite{2008ApJ...686..603J} found the average number of planets to be between 1.8 and 3.0 planets per system from a series of individual simulations with different initial planetesimal conditions, and that \citet{raymond2018solar} found the probability of forming planetary systems with a number of planets similar to our own to be $\sim$ 1\% based on dynamical arguments. Both results are very similar to our result but based on completely different and independent arguments. \\

%--------------------------------------------------------------------
\section{Discussion} \label{sec:Dis}

We find an anti-correlation between orbital eccentricity and multiplicity of known exoplanet systems similar to the reports by previous studies \citep{2015PNAS..112...20L,2017A&A...605L...4Z}. Our planet sample and method differ from the investigation by \citet{2015PNAS..112...20L} by including planets discovered by all detection methods, not just RV, and from both studies by including a much larger dataset and by comparing the results obtained based on different databases with different selection criteria. In addition we have chosen to consider systems as units unlike the previous studies, that treated each planet separately. 
When comparing our investigation to the previous ones, it should be noted that we, of course, share a great part of our data sample, and although the larger dataset in our analysis has allowed for a more restrictive debias process, all our analyses are biased by the basic limitation by the RV technique (biased towards larger planets), and the transit technique (biased towards ultra small orbits).

The fact that we include all planets regardless of detection methods has shown us that similar eccentricity-multiplicity correlations can be found for the full sample, and RV- and transit subsamples respectively, though with slightly different fits as discussed above. Explicitly, we also studied the eccentricity-multiplicity correlation for subsamples of hot-Jupiters, cold-Jupiters, and super-Earths separately, and found that also these subsamples followed the same general tendency. This shows that the correlation is not solely caused by the giant planets, that currently dominate our observations, or by planets at very short periods, but might also apply to the population of planets that have yet to be observed, with smaller masses and at larger periods.\\

A correlation between orbit eccentricity and multiplicity is supported by several other studies.
Surveys conducted by \citet{howard2013observed} and \citet{wright2009ten} found lower orbit eccentricities among planets in multi-planet systems compared to single planets. \citet{howard2013observed} suggests that the trend could be due to severe planet-planet scattering in the existing single-planet systems where giant planets have excited the eccentricities of its previous companions before ejecting them. Multi-planet systems have had fewer scattering events (otherwise they would no longer be multi-planet), and have thereby been allowed to stay in stable low-eccentricity orbits. \citet{wright2009ten} argues that multi-planet systems will naturally favour low-eccentricity orbits because of the need for high orbital stability in the system. The stability of multi-planet systems was studied further by \cite{2017AJ....153..210H}, who found that a single outer high-eccentricity giant planet would greatly affect the stability of an inner system, by reducing the multiplicity and exciting the eccentricities of the remaining planets.
Both \citet{wright2009ten} and \citet{2007DDA....38.1501C} support the theory of single high-eccentric planets as a  result of ejecting companions. The ejection of planets from planetary systems have been confirmed by \citet{mroz2017no}, who from analysis of short duration events in 6 years of microlensing data have found free floating planets of both Jupiter- and Earth-size, although they also conclude that the abundance of free floating planets is small, and can therefore only account for the eccentricity of a small fraction of the M1 systems. A study by \cite{xie2016exoplanet} have also reported lower eccentricities in multi-planet systems. This study measured the eccentricity distribution of a sample of transit planets using transit duration statistics, and found that single-planets in general show eccentricities of $e\approx 0.3$, whereas their multi-planet counterparts have average eccentricities of $e\approx0.04$. \cite{xie2016exoplanet} found all planets from multi-planet systems to follow similar eccentricity distributions, and so, found no general correlation between eccentricity and multiplicity. \\
Several studies have suggested that the correlation between eccentricity and multiplicity originate in an underlying correlation between multiplicity on the stability of the system, or the angular momentum deficit (AMD) \citep{laskar2017amd,turrini2020normalized,he2020architectures}. In their study, \citet{he2020architectures} recreate the multiplicity distribution observed at the Kepler data, using a forward model, by looking at the AMD-stability limit. They find that the median eccentricities as a function of multiplicity follow a power law correlation for all multiplicites from one to ten. Their model predicts that intrinsic single-planet systems have higher eccentricities than multi-planet systems, whereas most observed single-planet systems contain yet undiscovered planets, similar to our assumptions in Sec. \ref{sec:meanM}. Like previous studies \citet{he2020architectures} argued that the correlation between intrinsic multiplicity and the eccentricity of the systems was caused by the fact that the AMD-stability criteria puts strong demands on the total system AMD and minimum system period ratio, in order for no planet orbits to cross, and thereby destabilizing the system.

The eccentricity-multiplicity anti-correlation is opposed by \citet{bryan2016statistics} and \citet{dong2013warm}, who found lower eccentricities among single planets compared to planets with outer companions. Both surveys mainly focus on jovian planets, Dong et. al. solely on warm-Jupiters with jovian companions. \cite{dong2013warm} suggest that their results indicate that planet-planet interactions are not the dominating mechanism for creating short-period jovian planets, as opposed to the suggestions by several other studies \citep{rasio1996dynamical,marzari2002eccentric,2007DDA....38.1501C,nagasawa2008formation}. 

As argued by \citet{bryan2016statistics} a significant uncertainty is involved with the investigation by \citet{2015PNAS..112...20L} and some of this apply to our study as well. Many planets included have small semi major axes (the majority within 1 AU), and the low eccentricities found in high-multiplicity systems might reflect the fact that systems this closely packed would not be able to remain stable at higher eccentricities. With our larger data sample we have found similar correlations between RV- and transit subsamples which lowers the probability that the correlation is caused by observational biases. \cite{bryan2016statistics} further emphasize the uncertainty related to the fact that \cite{2015PNAS..112...20L} do not account for the individual errors of each of the listed eccentricities, which could also pose an uncertainty for this study.

Since we have not included the listed uncertainties to the eccentricities of each individual planet, we have not accounted for the uncertainty involved with the estimation of orbit eccentricity of exoplanets. In addition to this, previous studies have found that many eccentricities are systematically overestimated \citep{shen2008eccentricity}, and that some seemingly high-eccentricity single planets can turn out to have an unknown outer companion that artificially increase their estimated eccentricity \citep{fischer2001planetary}. The latter, fits our eccentricity-multiplicity correlation, with a decrease in eccentricity for an increasing number of known planets, it does however represent an uncertainty to our calculated model.

Unlike this study, the study by \cite{2017A&A...605L...4Z} did account for the uncertainties related to the eccentricity measurements. When calculating the mean eccentricities, \cite{2017A&A...605L...4Z} weighted their data with one over the listed uncertainties, which resulted in a steeper curve for the power law correlation compared to unweighted data. Especially the M2 systems seemed to differ between the weighted and unweighted data by having a significantly higher mean eccentricity in the weighted sample. They did not give an explanation as to why the low-eccentricity M2 planets should have generally higher uncertainties. In this study we find that the M2 systems have eccentricities that fit the general eccentricity multiplicity correlation for $M > 1$  without correction for the uncertainties. In our analysis only the M1 systems falls substantially below the power law fit, but since no M1 systems were included in the analysis by \cite{2017A&A...605L...4Z} we are not able to compare this trend to their results.

\section{Conclusion} \label{sec:Con}
During this study we have investigated the correlation between orbital eccentricity and multiplicity for 1171 planets distributed in 895 systems listed in the database \textit{exoplanet.eu}. 
We found a strong correlation between average eccentricity and multiplicity for all systems with two or more planets, which could be expressed as $e(M)=0.429\cdot M^{-0.93}$ (Eq.\,\ref{eq:e(M)}). The Solar System fits this trend, without being included in the making of the power law, whereas the average eccentricity of the observed M1 systems were markedly lower than predicted from Eq.\,\ref{eq:e(M)}. It is not unexpected from standard core accretion theory that the M2 to M$_{\rm max}$ systems fit the same power law distribution, but it  is surprising that the M1 systems fall substantially below the correlation. 
The eccentricity-multiplicity correlation is investigated for at number of different subsamples, in order to explore the stability of the power law correlation, and investigate possible explanations for the deviating M1 average eccentricity. All subsamples show the same general pattern, with all multiplicities fitting a power law correlation well, except the M1 systems having consistently lower eccentricities. The analyzed subsamples include:  different planet populations (divided into hot-Jupiter-, cold-Jupiter-, and super-Earth systems), planets detected by the RV- or transit method respectively, etc.

In order to investigate some of the implications of the power law trend, we speculated on the potential consequences, if the trend that was found for $M>1$, in reality applies to all multiplicities. Following the idea that Eq. \ref{eq:e(M)} describes the true eccentricity-multiplicity correlation, we assumed that the seemingly low eccentricities of the M1 systems were caused by a combination of some systems having been circularized through planet-star interactions, and others containing yet undiscovered planets. Correcting for these assumptions, a probability distribution over the different multiplicities was expressed by Eq.\, \ref{eq:prob2}, and based on this the mean multiplicity among the observed systems was estimated to $\langle M \rangle \approx 2.5$, while the probability of a system having eight planets was $\sim 1\%$. 

It is not surprising that the probability of finding high-multiplicity systems comes out this low, after all there are very few known exoplanetary systems with more than 6 planets, but it is assuring that the average number of planets in a "standard" exoplanet system in our Galaxy comes out very close to the number predicted independently from numerical simulations of planetesimal collisions (\cite{2008ApJ...686..603J}) and that the probability of finding Solar System like multi-planet systems comes out close to recent independent predictions from dynamical simulations \citep{raymond2018solar}.
This indicates that the orbit eccentricities of the Solar System planets are not unusually low, when the multiplicity of the system is taking into account, but rather that the number of planets in our Solar System is unusually high.The rarity of the large number of planets in our Solar System, and the corresponding low value of the orbital eccentricities, raise the simple and central, but speculative, question “Is there a connection between the high number of planets in our Solar System and the fact that we are here?”.\\ 

\section*{Acknowledgments}
This research has made use of The Extrasolar Planets Encyclopaedia at \textit{exoplanet.eu} and the Exoplanet Orbit Database and the Exoplanet Data Explorer at \textit{exoplanets.org} .
We are thankful for clarifying discussions with F. Roques about the selection criteria used by \textit{exoplanet.eu} and with J. Wright about the selection criteria used by \textit{exoplanets.org}. We acknowledge funding from the European Union H2020-MSCA-ITN-2019 under Grant no. 860470 (CHAMELEON) and from the Novo Nordisk Foundation Interdisciplinary Synergy Program grant no. NNF19OC0057374.
We are grateful to an anonymous referee, whose valuable input improved the analyses and argumentation throughout the paper. 

%%%%%%%%%%%%%%%%%%%% Data Availability %%%%%%%%%%%%%%%%%%

\section*{Data Availability}
The data underlying this article are available in The Extrasolar Planets Encyclopaedia, at \url{http://exoplanet.eu/catalog/}.



\bibliographystyle{mnras}
\bibliography{ref} % if your bibtex file is called example.bib

% Don't change these lines
\bsp	% typesetting comment
\label{lastpage}
\end{document}